\documentclass[10pt,aps,twocolumn,preprintnumbers, ,nofootinbib]{revtex4-1}
\usepackage[T1]{fontenc}
\usepackage{booktabs}
\usepackage[compat=1.1.0]{tikz-feynman}
\usepackage{feynmp}
\usepackage{xcolor}
\usepackage[utf8]{inputenc}
\usepackage{graphicx}
\usepackage{float}
\usepackage{mathtools}
\usepackage{amsmath}
\usepackage{array}
\usepackage{amsmath}
\usepackage{graphicx}
\usepackage{dcolumn}
\usepackage{amssymb}
\usepackage{balance}
\newcommand{\be}{\begin{equation}}
\newcommand{\ee}{\end{equation}}

\usepackage{graphics}
\usepackage{bm}        
\usepackage{amssymb}   
\usepackage{xcolor}
\usepackage{graphicx}
\usepackage{subfigure}
\usepackage[colorlinks=true,urlcolor=blue,linkcolor=black]{hyperref}

\usepackage{array}
\usepackage{mathtools,slashed}
\usepackage{multirow}
\usepackage{rotating}

\usepackage{array}
\usepackage{mathtools}
\usepackage{multirow}
\usepackage{rotating}
\usepackage{amsmath}
\usepackage{xcolor}
\usepackage[T1]{fontenc} 

\usepackage{tikz}

\usepackage{upgreek}

\usepackage{extarrows} 
\usepackage{amsmath}
\usepackage{graphicx}
\usepackage{dcolumn}
\usepackage{amssymb}
\usepackage{bm}
\usepackage{wasysym}
\usepackage{verbatim}
\usepackage{mathtools,slashed}

\newcommand{\p}{\partial}

\newcommand{\an}{\quad \textmd{and} \quad }

\newcommand{\bt}{\boldsymbol{\tau}}

\newcommand{\bT}{\boldsymbol{\tau}}

\newcommand{\bea}{\begin{eqnarray}}
\newcommand{\eea}{\end{eqnarray}}

\newcommand{\bp}{{\boldsymbol{\pi}}}
\newcommand{\bS}{{\boldsymbol{\Sigma}}}

\definecolor{mygray}{gray}{0.5}

\definecolor{mypink}{rgb}{0.03, 0.8, 0.1}

\begin{document}

\preprint{CERN-TH-2022-086}

\title{Ultra-Light Pion and Baryon WIMPzilla Dark Matter}

\author{Azadeh Maleknejad}

\affiliation{Theoretical Physics Department, CERN, 1211 Geneva 23, Switzerland}

\author{Evan McDonough}

\affiliation{Department of Physics, University of Winnipeg, Winnipeg MB, R3B 2E9, Canada}


\begin{abstract}
We consider a dark confining gauge theory with millicharged Ultra-Light Pions (ULP) and heavy baryons as dark matter candidates.  The model simultaneously realizes the ultra-light (STrongly-interacting Ultralight Millicharged Particle or ``STUMP'') and superheavy (``WIMPzilla'') dark matter paradigms, connected by the confinement scale of the dark QCD.  It is a realization of millicharged ULDM, unlike conventional axions, and exhibits a mass splitting between the charged and neutral pions.   ULPs can easily provide the observed density of the dark matter, and be cosmologically stable, for a broad range of dark QCD scales and quark masses. The dark baryons, produced via gravitational particle production or via freeze-in, provide an additional contribution to the dark matter density. Dark matter halos and boson stars in this context are generically an admixture of the three pions and heavy baryons, leading to a diversity of density profiles. That opens up the accessible parameter space of the model compared with the standard millicharged DM scenarios and can be probed by future experiments. We briefly discuss additional interesting phenomenology,  such as ULP electrodynamics,  and Cosmic ULP Backgrounds.
 \end{abstract}

\maketitle
\flushbottom

\tableofcontents




\section{Introduction}


 Few things are known with the degree of certainty that it is known dark matter exists: Cosmic Microwave Background (CMB) data, in the context of the $\Lambda$CDM cosmological model, indicates a non-zero abundance of dark matter to a statistical significance greater than 70$\sigma$ \cite{Planck:2018vyg}. However, beyond its gravitational influence, little is known about the identity of dark matter. Emblematic of our ignorance of dark matter is the mass of its constituent degrees of freedom, which could range from $10^{-22}$ eV to $10^{15}$ grams. 

Dark matter candidates in the sub-eV mass range are collectively known as Ultra-Light Dark Matter (ULDM) \cite{Ferreira_2020}. In the decades since the advent of axion dark matter \cite{Preskill:1982cy,Abbott:1982af,Dine:1982ah}, this class of models has been populated with Fuzzy \cite{Hu:2000ke,Hui:2016ltb}, Bose-Einstein condensate \cite{Boehmer:2007um}, bosonic superfluid, \cite{Berezhiani_2015,Ferreira:2018wup} fermionic superfluid \cite{Alexander:2016glq,Alexander:2018fjp}, and superconducting, e.g., STrongly-interacting Ultralight Millicharged Particle (STUMP) \cite{Alexander:2020wpm}, dark matter candidates, amongst others. Axion dark matter is but the tip of the ULDM iceberg.

\begin{figure}[h!]
\begin{center}\label{Setup-}
\includegraphics[width=0.48 \textwidth]{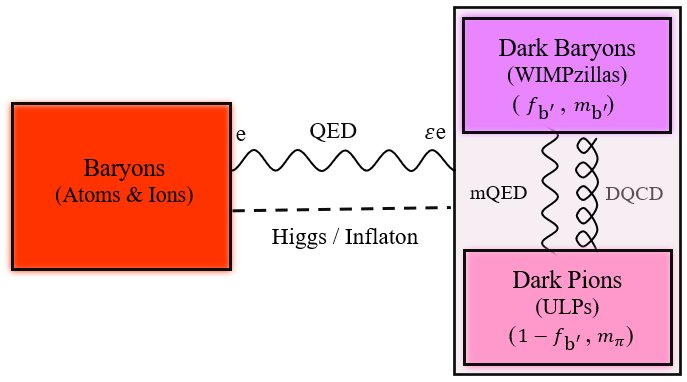} 
\caption{\label{fig:setup} Taxonomy of the ULP-WIMPzilla model. Dark matter is comprised of ultra-light pions (ULPs) and superheavy dark baryons, with masses $m_{\pi}$ and $m_{b}$ jointly determined by the confinement scale and quark mass, constituting a fraction of the dark matter $f_{b'}$ and $f_{\pi}=1-f_{b'}$ respectively. The ULPS and WIMPzillas interact with each other through the dark strong force (DQCD) and visible electromagnetism by their millicharges $\varepsilon e$ (mQED). They also interact with the Standard Model electromagnetically as well as gravitationally, through the Higgs, or via the inflaton.}
\end{center}
\vspace{-0.5cm}
\end{figure}

These models are in part compelling due to their potential for resolving small-scale tensions in the $\Lambda$CDM model, such as the core-cusp problem, missing satellites, and galactic rotation curves \cite{Bullock_2017, Chen_2017, Berezhiani_2018, Niemeyer_2019}.  However the modern science case is broader and deeper than this alone (see \cite{Antypas:2022asj} for a recent overview). ULDM exhibit a wide array of interesting phenomenology, such as vortices \cite{Rindler_Daller_2012,Schobesberger:2021ghi,Berezhiani_2021, Hui_2021} and their imprints \cite{Alexander:2021zhx,Alexander:2019puy}, electromagnetic signatures (e.g., \cite{Sikivie:2013laa}, see \cite{Zyla:2020zbs} for a review), gravitational waves \cite{Croon:2018ftb,Croon:2018ybs,Kitajima:2020rpm,Machado:2019xuc,Yuan:2021ebu}, and in varied particle physics contexts, such as muon $g-2$ experiments \cite{Janish:2020knz} and neutrino oscillation experiments \cite{Dev:2020kgz}. This wide array of potential observables for both current and next generation experimental efforts motivates a thorough study of the ULDM model space.

\begin{figure}[h!]
\begin{center}
\includegraphics[width=0.49 \textwidth]{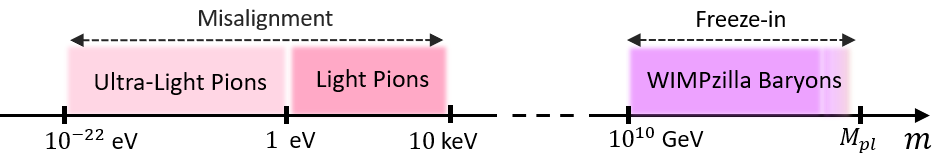} 
\caption{\label{fig:mass-spectrum-I}  Summary of dark matter production mechanisms with respect to mass. This setup predicts an admixture of light dark pions  (Sec. \ref{MA-P}) and superheavy baryons (Sec. \ref{FI-B}) as the cold dark matter. Light and ultra-light pions are produced via the misalignment mechanics in the early universe. The baryon WIMPzillas can be generated gravitationally or via freeze-in mechanism by an inflaton portal, Higgs portal, and millicharged QED interactions. }
\end{center}
\vspace{-0.5cm}
\end{figure}

In this paper we study the emergence of millicharged ULDM in the context of a confining gauge theory. We find that the pions of this theory, namely, ultra-light pions (ULPs), are an excellent dark matter candidate. These are analogous to the Standard Model pions in their field theory formulation, but share important properties with the conventional axion, such as the cosmological evolution as a coherent field. ULPs are distinguished from a conventional axion in part by their characteristic spectrum: a neutral pion $\pi^0$ and charged pions $\pi^\pm$. The masses of the pions are dictated by the confinement scale, the quark mass, and the charges of the quarks. As a concrete example, we consider the possibility that the dark quarks have a small electric charge. While so-called millicharged dark matter has been widely explored, this is the first electrically millicharged ULDM candidate (for 1 eV to 10 keV millicharged DM models see \cite{Bogorad:2021uew}). ULPS are a close cousin of the STUMP model proposed in \cite{Alexander:2020wpm}, and provide an alternative model realization of STUMPs.


The ULPs themselves are but the tip of another iceberg, comprised of the mesonic and hadronic states of the theory. For concreteness we focus on a dark sector similar to Quantum Chromo-Dyamics (QCD), namely we consider an SU(3) gauge theory with two flavours of light  quark, which we refer to as dark QCD (see Fig. \ref{fig:setup}).  For past works on dark QCD, see, e.g. \cite{Garani:2021zrr, Tsai:2020vpi, Dondi:2019olm}. In our setup, the lightest baryons, namely a dark proton and dark neutron, are stable and themselves can be excellent dark matter candidates.  The dark baryons are naturally in the realm of WIMPzillas \cite{Chung:1998rq,Chung:1998ua,Kuzmin:1998kk,Chung:2001cb}, also known as superheavy dark matter, since the dark baryon mass is anchored to the confinement scale, which is in turn related to the pion decay constant, which is $\gtrsim 10^{10} {\rm GeV}$ in order to realize ULPs as a dark matter candidate. See Fig. \ref{fig:mass-spectrum-I}.

Superheavy dark matter has its own rich phenomenology \cite{Carney:2022gse}, and disentangling superheavy DM candidates, e.g., spin-0, spin-1/2, spin-1 
\cite{Kolb:2020fwh}, spin-3/2 \cite{Kolb:2021xfn}, and spin-$s$ \cite{Alexander:2020gmv}, is an interesting direction for future work. In the ULP model, the superheavy dark matter is a fermion, and itself a composite state.  This setup has a very rich phenomenology which is specified in terms of three unknown scales, i.e. mass of the dark quarks, scale of the dark confinement, and scale of inflation. The focus of current work is at the limit where the dark confinement happens before inflation. The exhaustive study of the setup throughout the parameters will be presented in \cite{longer-v}.

The outline of this paper is as follows: in Sec.~\ref{sec:EFT} we introduce our dark QCD theory and the spectrum of particles in the confined phase. in Sec.~\ref{sec:DM} we develop ULPs as a dark matter candidate, along with their superheavy dark baryons. In Sec.~\ref{sec:halos} we find approximate solutions to ULP boson stars and Fuzzy ULP halos. We conclude in Sec.~\ref{sec:discussion} with a discussion of other directions for ULP detection, in particular, by utilizing their interactions with the photon. The details of the calculation of the number density of heavy dark baryons is presented in Appendix \ref{Appx}.

{\bf{Notation:}}  the dark $SU(N)_x$ gauge field and its quarks are denoted as $\bf{X}_{\mu}$ and $\boldsymbol{\chi}$ respectively. The SM Higgs doublet is denoted as $\rm H$ and $B_{\mu}$ is the SM's $U(1)_Y$ hypercharge. Throughout this work, unless otherwise specified, by ``quark'', ``pion'', ``baryon'', ``eta-prime'', and ``glueball'', refer to dark sector components.  Here $(\pi^0,\pi^+,\pi^-)$ are the dark pions, $\eta'$ is the dark eta-prime, and $(b,n)$ denote the dark baryons. The dark baryon number is shown as $\rm B'$. The Hubble parameter is shown as $H$.  Finally, to avoid confusion with the SM baryon density, we used $\Omega_{b'}$, and $f_{b'}$ for the relic density and fraction of energy in the dark baryons.



\section{Effective Field Theory and Cosmological History of Dark QCD}
\label{sec:EFT}


We consider a confining dark gauge symmetry $SU(N)_x$ coupled to two dark quarks ( ``up'' and ``down'') in the fundamental representation of the dark color $SU(N)_x$ as
\bea \label{theory}
\mathcal{L}_{X} = i \bar{\boldsymbol{\chi}} \slashed{D} \boldsymbol{\chi} - \boldsymbol{m}_{\chi} \bar{\boldsymbol{\chi}} \boldsymbol{\chi} - \frac12 \rm{Tr}{\bf{X}}_{\mu\nu}{\bf{X}}^{\mu\nu} ,
\eea
where $\slashed{D}= \slashed{\nabla} - i g_x \slashed{\bf{X}},$
in which 
 $\boldsymbol{m}_{
\chi} = {\rm{diag}}(m_u,m_d) $ is a quark mass matrix. 
The quarks $\boldsymbol\chi_{u,d}$ carry dark baryon numbers as $b'_u=b'_d=\frac13$.  
We are interested in the limit that $m_{\chi}$ is negligible compared to the confinement scale of the $SU(N)_x$, i.e., $m_{\chi} \ll \Lambda_{x}$. The theory Eq.~\eqref{theory} then has an approximate $SU(2)_L \times SU(2)_R \times U(1)_{V} \times U(1)_A$ chiral symmetry, under which ${\boldsymbol{\chi}}_{L,R}$ transform as a doublet.

At energy scales below the confinement scale of $SU(N)_x$, 
the dark QCD (``DQCD'') vacuum spontaneously breaks the $SU(2)_L\times SU(2)_R$ 
global symmetry down to the diagonal subgroup $SU(2)_V$, via the x-quark-antiquark condensate as
\bea\label{V}
\langle \bar{\boldsymbol{\chi}}_i^L \boldsymbol{\chi}_j^R \rangle = V^3 \delta_{ij},  \quad i,j \in \{u,d \},
\eea
where $V \sim \Lambda^3_{x}$. The quarks are then confined into mesons and hadrons.  We have two composite pseudo-scalar states,
\bea
\boldsymbol{\pi} = \pi^a \bT_a, \quad \eta',
\eea
where  $\bT_a$ are the generators of the $SU(2)$ algebra, the $SU(2)$ triplet $\boldsymbol{\pi}$ are Nambu–Goldstone bosons (NGBs) corresponding to spontaneous symmetry breaking of $SU(2)_A$, and $\eta'$ is a massive singlet, which are all their own antiparticles. There is additionally a spectrum of mesons and baryons, just as in standard visible QCD. However, unlike QCD, here there is no analog of the electroweak interactions, which makes the pions stable.

\begin{figure}[h!]
\vspace{-0.1cm}
\begin{center}
\includegraphics[width=0.48 \textwidth]{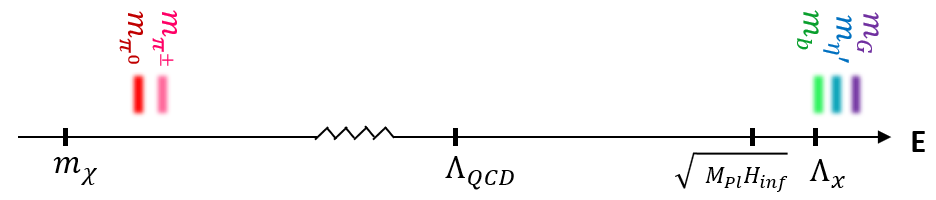} 
\caption{\label{fig:mass-spectrum-II} Mass spectrum of the theory. The dark baryon and $\eta'$ masses are around $\Lambda_x$ and the lightest scalar glueball mass is $m_G\approx 6 \Lambda_x$. The neutral pion mass is $m_{\pi^0}\simeq \sqrt{m_\chi \Lambda_x}$ that for very light quarks $m_{\pi^0}\ll \Lambda_x$. The mass splitting between charged and neutral pions is $\frac{m_{\pi^{\pm}}^2  -m_{\pi^{0}}^2 }{m_{\pi^{0}}^2 } \sim   e^2 \frac{ \varepsilon^2  F_{\pi}}{m_u+m_d}$.  }
\end{center}
\vspace{-1cm}
\end{figure}

The mass spectrum of the lowest-lying mesonic and hadronic states is shown in Fig.~\ref{fig:mass-spectrum-II}.
Before studying the spectrum of the theory in detail, we first consider ways in which dark QCD may be coupled to the visible Standard Model.

\subsection{Portals to the Standard Model}

Any dark sector is unavoidably coupled to the SM gravitationally. Besides, 
there are numerous portals of dark QCD to the standard model, see., e.g. \cite{Cline:2021itd}. For example, the quarks of the dark sector might be connected to the SM via a small ``milli-'' charge under SM hypercharge, or through a coupling to the SM Higgs. In particular, the dark quarks may have a very small hypercharge, i.e.,
\be
\slashed{D}= \slashed{\nabla}- i g_x \slashed{\bf{X}} - i \varepsilon_{u,d} e \slashed{B},
\ee
where $e$ is the electric charge, $B_{\mu}$ is the SM hypercharge and $\varepsilon_{u,d}\ll1$ are two very small numbers. For millicharged dark sectors as extensions of the Standard Model embedded in unifying gauge framework see \cite{Feldman:2007wj}.

\begin{figure}[h!]
\begin{center}
\includegraphics[width=0.3 \textwidth]{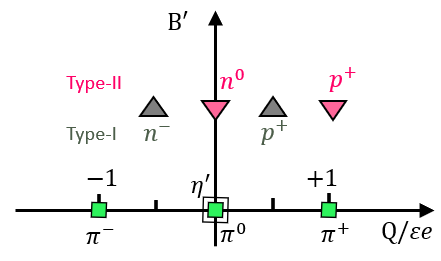} 
\caption{
\label{ch-b} Composite states in charge-baryon number plane. Based on the charge type of the dark quarks in Table \ref{tab:milli} dark baryons have different charges. In type-I, neutron and proton have $\pm\frac12\varepsilon e$ electric charges respectively while in type-II neutron is neutral and proton have $+\varepsilon e$ charge. }
\end{center}
\end{figure}

To this end we consider two types of scenarios as shown in Table \ref{tab:milli}. We label as Type I the case in which the up and down quarks have equal but opposite hypercharge, and label as Type II the case in which, analogous to the SM quarks, the dark quarks have charge $+(2/3) \epsilon e$ and $-(1/3) \epsilon e$. In both cases, there is an electrically-neutral pion and two electrically-charged pions $\pi^{\pm}$ with charge $\pm \epsilon e$. In Type I there is a electrically-neutral neutron and ellectrically-charged proton, while in Type II both the neutron and proton are charged. (See Fig. \ref{ch-b}).

\begin{table}[h]
    \centering
    \begin{tabular}{|c | c |c  |}
    \hline
     ~    &  $\varepsilon_u$ & $\varepsilon_d$  \\ \hline
     Type-I    & $+\frac12\varepsilon$   &  $-\frac12\varepsilon$  \\
      ~Type-II   & $+\frac23\varepsilon$    &  $-\frac13\varepsilon$   \\ \hline
    \end{tabular}
    \caption{We consider two types of millicharged dark quarks parametrized in terms of the dimensionless parameter $\varepsilon \ll 1$.}
    \label{tab:milli}
\end{table}

A second possibility is a Higgs portal to the Standard Model. For example, the dark quarks can be directly coupled to the SM Higgs via a dimension-5 operator $\mathcal{L}_{{\bf \chi- \rm{H}}} = \sum_{i=u,d} \frac{y_{i}}{\Lambda_{ \rm H}}  {\bf \rm H}^{\dag} {\bf \rm H} ~ \bar{\boldsymbol{\chi}_i}\boldsymbol{\chi}_i$,
where ${\bf \rm H}$ is the SM Higgs doublet, and $\Lambda_{\bf \rm H}$ is the mass of the heavy mediator between DM and Higgs. In the confining phase, this interaction generates effective couplings of the Higgs to both the mesons and the baryons. A variation on this is to consider portals of the confined phase, and in particular consider  {\it dark-baryon-philic} couplings of the Standard Model, analogous to leptophilic dark matter \cite{Fox:2008kb} and leptophobic dark matter \cite{FileviezPerez:2018jmr}. Dark-baryon-philic interactions can emerge from simple UV completions, e.g., in models such as \cite{Duerr:2013dza,Ma:2020quj} wherein the global $U(1)$ dark baryon number is promoted to a spontaneously broken gauge symmetry, as proposed in \cite{Alexander:2020wpm}. 

An additional possibility is to couple the dark sector to the SM indirectly, i.e. through an inflaton portal. This could again take the form of a baryon-philic coupling, or couplings to the full spectrum of the confined phase.

\subsection{The Spectrum}

We now consider the spectrum of composite states in the confining phase of dark QCD. Among the composite states, we will see that the pions and baryons are stable while $\eta'$ and glueballs are unstable and short-lived.

\subsubsection{Pions}

The dynamics of the pions in the confined phase is well described by chiral perturbation theory \cite{Scherer:2005ri}. In this framework, confinement can be described as a spontaneous symmetry breaking of a composite field $\Sigma_{ij}$, with additional terms which explicitly break the $SU(2)_A$  global symmetry. The action can be expressed as,
\be\label{Sigma-action}
S_{\Sigma} = \int dx^4 \sqrt{-g}  \bigg({\rm{Tr}}[ D_{\mu} \bS]^2  + m^2 \lvert  \bS \rvert^2 - \frac{\lambda}{4} \lvert  \bS \rvert^4\bigg) + {\cal L}_{\chi {\rm PT}},
\ee
where $D_{\mu} \bS$ is the covariant derivative defined as
\be
D_{\mu} \bS = \p_{\mu} \bS - i e A_{\mu} (\boldsymbol{Q} \bS - \bS \boldsymbol{Q}), \quad 
\boldsymbol{Q} = \textmd{diag}(
\varepsilon_u , \varepsilon_d), 
\ee
and ${\cal L}_{\chi {\rm PT}} $ denotes terms that explicitly break the global chiral symmetry.

In the symmetry-broken phase, $\Sigma_{ij}$ can be expanded around its vacuum as 
\bea
\bS_{ij} = \frac{F_{\pi}+\sigma(x)}{\sqrt{2}} \exp\big(\frac{2i\pi^a(x)\bt_a}{F_{\pi}} \big),
\eea
where $\langle  \bS \rangle = \frac{F_{\pi}}{\sqrt{2}} \textmd{diag}(1 , 1)$ is the vacuum solution. The effective action of the pions is then determined by ${\cal L}_{\chi {\rm PT}}$, which, at leading order,  is given by,
\be
{\cal L}_{\chi {\rm PT}}=
{\rm{Tr}}[   \bS^{\dag}\boldsymbol{m}_{
\chi}  + \boldsymbol{m}_{
\chi}^{\dag}\bS] + ... .
\ee
Moreover, the most generic $\chi$PT can have a term as \cite{Borasoy:2001ik, Kaiser:2000gs}
\be
{\cal L}'_{\chi {\rm PT}}= i f(\eta') 
{\rm{Tr}}[   \bS^{\dag}\boldsymbol{m}_{
\chi}  - \boldsymbol{m}_{
\chi}^{\dag}\bS] ,
\ee
where $f(\eta') $ is an odd function of $\eta'$. 
Although $\eta'$ is not a true Goldstone boson
due to the axial $U(1)_A$ anomaly of the strong interactions, it combines with the pions via the above effective interaction.

In the case of neutral and degenerate-mass quarks, the effective action of the pions takes an exceptionally simple form as
\be
\mathcal{L}_{\pi}\big\rvert_{_{m_u=m_d}} =   \frac12\bigg({\rm Tr} \p_\mu\bp \p^\mu\bp  + m_{u}V^3 {\rm Tr} \cos \left( \frac{\bp}{F_{\pi}}\right)  \bigg),
\ee
in strong resemblance to a conventional axion. 

More generally, expanding to quadratic order in the pions, we find
\bea
\mathcal{L}_{\pi} &=&   \frac12\p_\mu\pi^0 \p^\mu\pi^0 + D_\mu\pi^+ D^\mu\pi^- - \frac{m_{\pi^{0}}^2}{2} {\pi^0}^{2} - m_{\pi^{\pm}}^2 \pi^{+}\pi^{-} \nonumber\\ &+&\mathcal{L}_{\pi\gamma},
\label{eq:Lpi}
\eea
where the  mass of pions  is given by the Gell-Mann-Oakes-Renner relation
\be
m_{\pi^0}^2 = \frac{V^3}{F_{\pi}^2}(m_u+m_d),
\ee
for the neutral pion, where $F_{\pi}\sim V \sim \Lambda_x$, and 
\be \label{QED-mass}
m_{\pi^{\pm}}^2 = m_{\pi^0}^2 + 2\xi e^2 F_{\pi}^2 (\varepsilon_u-\varepsilon_d)^2,
\ee
for the charged pions, where the last term is the electromagnetic contribution to the mass splitting of $\pi^{\pm}-\pi^{0}$, and $\xi$ is an order one parameter which should be fit by data. Therefore, given that $\varepsilon_u \sim -\varepsilon_d= \varepsilon$, $\xi=\mathcal{O}(1)$ and $F_\pi \sim V$,  we have the mass splitting
\bea
\frac{m_{\pi^{\pm}}^2  -m_{\pi^{0}}^2 }{m_{\pi^{0}}^2 } \sim   e^2 \frac{ \varepsilon^2  F_{\pi}}{m_u+m_d},
\eea
which, depending on $\epsilon$, $F_{\pi}$, and the quark mass $m_\chi$, can range from negligibly small to very large. 

The pions are also coupled to the SM photon. The pion-photon interaction $\mathcal{L}_{\pi\gamma}$ in Eq.~\eqref{eq:Lpi} is the chiral anomaly of the $U(1)_A$ via the triangle diagram of the massive charged baryons as \cite{Schwartz:2013pla}
\be\label{L-gamma-pi0}
\mathcal{L}_{\pi\gamma} = \sum_{b^i} Q_{b_i}^2 \frac{N_c }{8 \pi^2 F_{\pi}} \pi^0 F\tilde{F},
\ee
where $N_c$ is the number of colors and $\sum_{b_i} Q_{b_i}^2=A \varepsilon^2 e^2$ is the charge of baryons inside the triangle loop where $A=1/2$ ( $A=1$ ) in type-I (type-II).
The pions also inherit quartic self-interactions and couplings coming both from the expansion of the cosine potential and from higher-order terms in chiral perturbation theory.

The coupling to the photon allows the neutral pion to decay as $\pi^0\rightarrow \gamma\gamma$. Taking the $m_{\pi^0} \ll m_{b}$ limit, this decay rate can be written as
\bea
\Gamma(\pi^0\rightarrow \gamma\gamma) = \frac{A \alpha^2_{e}}{64\pi^3} \frac{\varepsilon^4 \lambda^2 m_{\pi^0}^3}{m_{b}^2} = \frac{A\alpha^2_{e}}{64\pi^3} \frac{\varepsilon^4  m_{\pi^0}^3}{F_{\pi}^2}.
\eea
 Therefore, the lifetime of $\pi^0$ in type-I is twice the lifetime of $\pi^0$ in type-II. Assuming $m_u\sim m_d = m_\chi$ and given that $m^2_{\pi^0}\sim m_\chi \Lambda_x$, we find
\bea
&& \Gamma(\pi^0\rightarrow \gamma\gamma) \sim  \frac{2A\alpha^2_{e}}{64\pi^3} \varepsilon^4 m_\chi \bigg(\frac{m_\chi}{\Lambda_x}\bigg)^{\frac12} . 
\eea
In order for the $\pi^0$ to be long lived enough and a DM candidate, its lifetime should be longer that the age of Universe, $\Gamma=\hbar/\tau_{\rm{U}}< \hbar/(4\times 10^{17} ~s)$. This can be translated to the condition,
\bea
 \bigg(\frac{\varepsilon}{10^{-8}}\bigg)^4 \bigg(\frac{100~{\rm MeV}}{\Lambda_x}\bigg)^{\frac12} \bigg(\frac{m_{\chi}}{0.1 ~{\rm MeV}}\bigg)^{\frac32} &<& \frac32 .
 \label{eq:pion-lifetime}
\eea
Therefore, the pions can be made stable through any or all of a high confinement scale $\Lambda$, a small quark mass $m_\chi$, or a small millicharge $\varepsilon$. \\

\subsubsection{$\eta'$ Meson}

Similar to SM QCD, the global $U(1)_A$ symmetry of our dark QCD is broken by instantons. The mode associated to the $U(1)_A$ is the $\eta'$ meson,
which receives its mass due to the explicit breaking of $U(1)_A$ by the $SU(N)_x$ instantons as
\be
V(\eta') = \Lambda^4_x \cos\big(\frac{\eta'}{F_{\eta'}}\big),
\ee
where $F_{\eta'} \simeq \Lambda_{x}$ is the decay constant of $\eta'$. This endows the $\eta'$ with a mass $m_{\eta'}\sim \Lambda_x$, which is parametrically heavier than the pions. The effective Lagrangian of the $\eta'$ can be written as
\bea
\mathcal{L}_{\eta'} &=& \frac12\p_{\mu}\eta' \p_{\mu}\eta' + \Lambda^4_x \cos\big(\frac{\eta'}{F_{\eta'}}\big)  
- \frac{\alpha_{\eta'\!\gamma}}{4\pi}\frac{\eta'}{F_{\eta'}} F_{\mu\nu}\tilde{F}^{\mu\nu} \nonumber\\
&+& \alpha_{\eta'\!\pi} \lvert m_u - m_d\rvert \eta' \pi^0 \vec{\pi}.\vec{\pi},
\eea
where $\alpha_{\eta'\!\pi}$ and $\alpha_{\eta'\!\gamma}$ are two constants that should be given by the simulation. Note that the last term is isospin symmetry violating and hence is proportional to $\lvert m_u - m_d\rvert$.   

The dark $\eta'$ is unstable, just as in SM QCD. The decay channels of the $\eta'$ are the decay to two photons, i.e. $\eta'\rightarrow \gamma \gamma$, as well as decay to three pions, i.e. $\eta'\rightarrow \pi^+ \pi^- \pi^0$ and $\eta'\rightarrow \pi^0 \pi^0 \pi^0$. The decay rate of $\eta'$ to two photons can be estimated in terms of the same process in the SM as
\bea
\Gamma(\eta'\rightarrow \gamma\gamma) &\approx & \varepsilon^4  \bigg( \frac{m_{\eta'}}{m_{\eta'_{\rm SM}}}\bigg)^3 \Gamma(\eta'_{\rm SM}\rightarrow \gamma\gamma).
\eea
The $\eta'_{\rm{SM}}\rightarrow \gamma \gamma$ is the dominate decay channel of $\eta'_{\rm{SM}}$ which specifies its life-time as short as $\tau_{\eta'_{\rm{SM}}} \approx 3\times 10^{-21}~s$. In the $\eta'$ case, we have
\bea
\Gamma(\eta'\rightarrow \gamma\gamma) & \approx & \frac{1}{10^{-2} s} \bigg(\frac{\varepsilon}{10^{-13}}\bigg)^4  \bigg( \frac{m_{\eta'}}{10^{10} {\rm GeV}}\bigg)^3 ,
\eea
making the $\eta'$ is unstable on cosmological time-scales. Meanwhile the decay rate of $\eta'$ to three pions is given as
\bea
\Gamma(\eta'\rightarrow \pi\pi\pi) = \frac{\alpha^2_{\eta'\pi}(m_u-m_d)^2}{64(2\pi)^3} m_{\eta'},
\eea
which can be cosmologically fast or slow, depending on the mass splitting of the quarks, the $\eta'$-pion coupling, and the mass of the $\eta'$. Given these two decay channels, the generic expectation is that any initial $\eta'$ population should be unstable and decay to photons and pions.

\subsubsection{Glueballs}

In addition to the mesons and baryons, one expects that gluons can also form colourless states, called glueballs \cite{Glueballs}. In that case, dark Glueballs (DG) can form below the confinement scale as pure gluonic states. The properties of the glueballs in Yang-Mills have been studied in lattice gauge theory \cite{Bali:1993fb, Athenodorou:2021qvs, Morningstar:1999rf, Chen:2005mg, Loan:2005ff}. Their spectrum can entirely be parameterised by the confinement
scale of the theory, or equivalently lightest glueball mass, $m_G \sim 6 \Lambda_x$. The full QCD (including quarks), however, is more complicated and an active experimental and theoretical area of research. The lightest DG (denoted as $G$) is a scalar with quantum numbers $J^{PC}=0^{++}$, and its effective Lagrangian can be written as
\be
\mathcal{L}_{G} = \frac12\p_{\mu}G \p_{\mu}G + \frac12 m_G^2 G^2 +f_3 G^3 + f_4 G^4,
\ee
where $m_G$ is the mass, $f_3$ and $f_4$ are the self-couplings. The glueballs can be stable or decay
through a variety of portals to the mesons. Given that $G$ is a singlet under chiral symmetry, the DG-meson interactions is \cite{Jin:2002up}
\be
\mathcal{L}_{G\bS} = (g_1 F_{\pi} G  + g_2 G^2) {\rm{Tr}}[\bS^{\dag}\bS],
\ee
which leads to a $G$ and pions interaction term as
\bea
\mathcal{L}_{G\bp} \simeq \frac{\theta m_G^2}{2F_{\pi}} G \bp^2,
\eea
where $\theta=\frac{g_1 F_{\pi}^2}{(m_G^2-m_{\sigma}^2)}$ which its value should be given by the data. Therefore, the scalar DG decays into to pions with the decay rate
\be
\Gamma(G\rightarrow \pi^a\pi^a) = \frac{3\theta^2}{32\pi} \bigg(\frac{m_G}{F_{\pi}}\bigg)^2 \sqrt{m^2_G-m_{\pi}^2}.
\ee
Assuming that the $\theta$ is ${\cal}O(1)$, we have
\be
\Gamma(G\rightarrow \pi^a\pi^a)  \sim 10 \Lambda_x,
\ee
which implies that DG are unstable and very short-lived.  As a result, the dark glueballs have no cosmological effect in this setup. For a similar framework in a different part of parameter space with cosmological relevant glueballs see \cite{Dondi:2019olm} and \cite{Asadi:2022vkc}. 

\subsubsection{Dark Baryons} 

The lightest dark baryons are $p=uud$ and $n=ddu$ with masses $m_{p}\simeq m_{n} \approx N_x\Lambda_x$. Both baryons are stable, even in the charged case, since there are no electrons or neutrinos in the model, and dark baryon number is conserved. The dark baryons therefore serve as a potential dark matter candidate. For $\Lambda_x \gg {\rm GeV}$, this falls in the category of superheavy dark matter (recently reviewed in \cite{Carney:2022gse}), and in particular, superheavy dark fermions. These baryons are feebly coupled to the SM and can only get generated via the freeze-in mechanism in the early universe, or else gravitationally. We defer a discussion of the primordial production of baryons to Sec.~\ref{sec:DM}.


\section{Dark Matter}
\label{sec:DM}


\begin{figure*}
\begin{center}
\includegraphics[width=0.325 \textwidth]{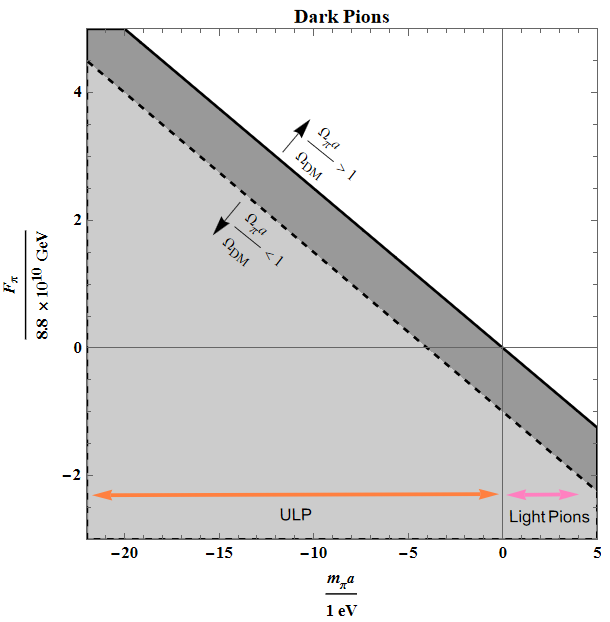}
\includegraphics[width=0.326 \textwidth]{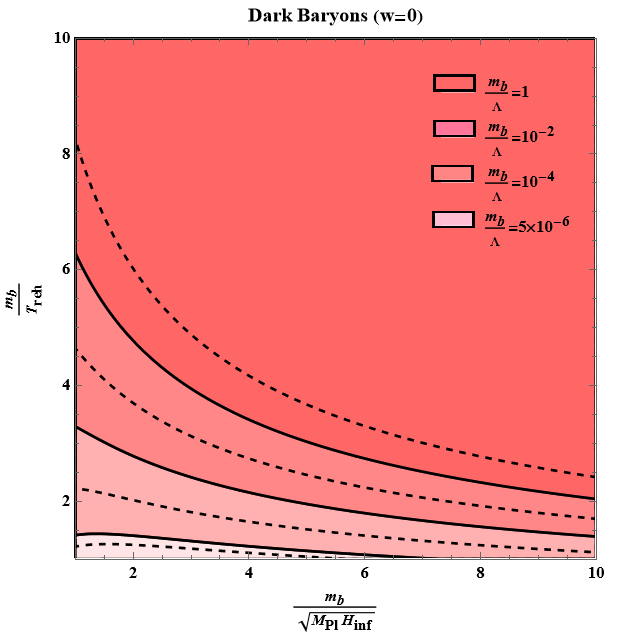}
\includegraphics[width=0.326 \textwidth]{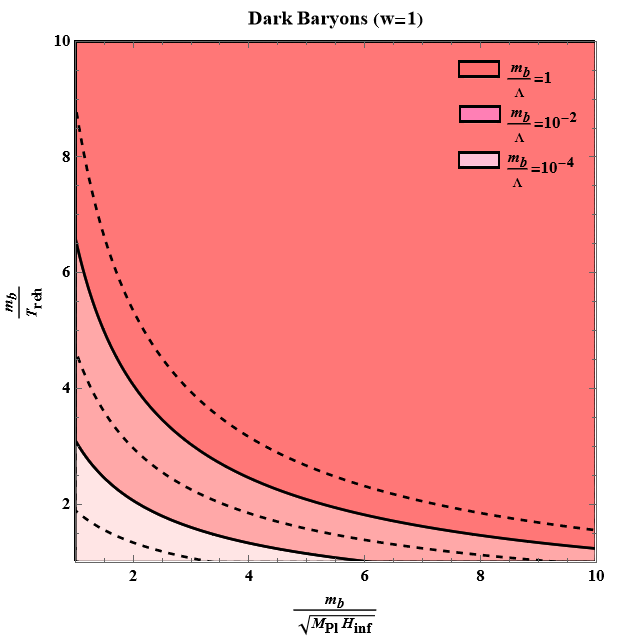} 
\caption{\label{fig:FI} The admixture of dark pions-WIMPzilla baryons with respect to parameters of the model, for freeze-in production of dark baryons (see Fig.~\ref{fig:GPP} for gravitational production). 
[{\it Left Panel}] Accessible parameter space for the pions produced by misalignment mechanism where the misalignment angle are $\theta_i\approx \pi$. The shaded region shows parameters corresponds to $\Omega_{\pi^a}/\Omega_{\rm DM}\lesssim 1$ and the dashed line marked shows $\Omega_{\pi^a}/\Omega_{\rm DM}= 10^{-2}$. [{\it Middle and Right Panels}] Freeze-in production of baryons (by Higgs portal Eq. \eqref{H-b}, Inflaton portal Eq. \eqref{Inf-b}, and QED portal Fig. \ref{Feyn}) for the preheating phase with $w=0$ (middle panel), and $w=1$ (right panel). The parameter $\Lambda$ denotes $\Lambda_{H}/y$ for the Higgs portal case, $\sqrt{2}\Lambda_{\phi}/\lambda$ in the inflaton portal case, and $13.7 m_b/\varepsilon$ for the freeze-in via millicharged case. 
Each shaded region represents part of the parameter space with $\Omega_{\rm b'}/\Omega_{\rm DM}\lesssim 1$ for the given value of $\frac{ m_{b}}{\Lambda}$. The dashed lines inside each shaded area makes $\Omega_{b}/\Omega_{\rm DM}= 10^{-2}$. }
\end{center}
\end{figure*}

We now focus our attention on the dark matter problem, namely, obtaining the observed relic density of cosmologically stable cold collisionless dark matter. As discussed above, the stable composite states in our model are the pions and the baryons. Here we consider these individually.

\subsection{Pion Dark Matter: Light and Ultra-Light}\label{MA-P}

As a simple cosmological history, we consider that confinement occurs at a high scale $\Lambda_x$, above the scale of cosmic inflation.  Similar to the conventional axion vacuum misalignment mechanism, the pion fields are each initialized with an initial value comparable to the pion decay constant. The phase of cosmic inflation serves to both homogenize the pion fields, and redshift away any thermal relics from the confining phase transition. 

In order to constitute the observed dark matter, we demand the dark pions satisfy the following requirements:
\begin{enumerate}
\item Ultra-Light and Light: We are interested in ultra-light pions (ULPs), $m_{\pi^a}< 1 \, \rm eV$, and light pions $m_{\pi^a} = (1 \, \rm eV-10 \, keV$). They are generated via the misalignment mechanism in the early universe.
\item Stability: In our dark QCD, with neutral quarks, the pions are stable, since there are no electroweak interactions. In the millicharged case, this requires eq.~\ref{eq:pion-lifetime} be satisfied.
\item Relic Density: The three pions together constitute the observed dark matter density. The present day pion abundances is given by,
\begin{equation}\label{misa}
    \Omega_{\pi} = \frac{1}{6} (9 \Omega_r)^{3/4} \frac{F_\pi^2}{M_{pl}^2} \displaystyle \sum _{a=0,\pm} \left(\frac{m_{\pi^a}}{H_0} \right)^{1/2}  \theta_{\pi^a} ^2
\end{equation}
in analogy to the usual axion case \cite{Marsh:2015xka}. 
\item Isocurvature Modes: Light pions pick up isocurvature perturbations during inflation which is strongly constrained by the CMB \cite{Planck:2018jri}. 
Assuming that $\theta_{\pi^a} \sim \pi$, this leads to an upper bound on the scale of inflation, as
\bea
 \frac{H_{inf}}{M_{pl}}\lesssim 8.8 \times 10^{-5} \frac{\Omega_{DM}}{\Omega_{\pi^a}} \frac{F_{\pi}}{M_{pl}} .
\eea
\end{enumerate}
Finally the requirement that the dark matter be collisionless is satisfied due to the fact that the pion self-interactions and couplings are suppressed by the decay constant $F_{\pi}$, and additionally by the requirement that the millicharge be small enough to extend the pion lifetime to greater than the age of the universe. In the following, we discuss the parameter space of dark pions. (See the left panel of Fig. \ref{fig:FI}).

\subsubsection{Ultra-Light Pions (ULPs)} 

 The dark pions with masses $m_\pi < 1 {\rm eV}$ evolve as a coherent scalar field. Similar to Axion-like particles (ALPs), this implies the ULPs evolve as cold pressureless dark matter in the late universe, when each ULP field oscillates in the minimum of its potential.  Demanding that the pions constitute an ${\cal O}(1)$ fraction of the observed dark matter density enforces a lower bound on the pion decay constant, and hence the dark QCD scale, given by $F_\pi > 8.8 \times 10^{10}$ GeV. From these considerations, and the relation $m_{\pi^0} ^2\sim m_\chi \Lambda_x$, we may deduce the mass range for the dark quarks, as
\begin{equation}\label{mass-ULP}
    m_\chi < 5 \times 10^{-20} {\rm eV}.
\end{equation}
Further demanding that the charged pions, with mass given by Eq. \eqref{QED-mass}, remain in the ultra-light regime, $m_{\pm} < $ eV,  puts an upper bound on the charge of the dark quarks as
\be
\varepsilon e < \frac{1~ {\rm eV}}{F_{\pi}}.
\ee
 Curiously, the mass bound Eq.~\eqref{mass-ULP}  is within a few orders of magnitude of the benchmark Fuzzy Dark Matter mass. 
For the purposes of this work, we do not consider the fine-tuning of the dark quark mass to be any more concerning of a problem then it already is for bosonic ultra-light DM (e.g. axion or fuzzy DM).

\subsubsection{Light Pions}

The light dark pions have masses in the interval $m_{\pi^a}= (1 \,{\rm eV} - 10 \, \rm{keV})$. Using $H_0\approx 2 \times 10^{-33}\, \rm{eV}$ in Eq. \eqref{misa}, we have
\be
\Omega_{\pi^a} \approx 0.25 \bigg( \frac{F_{\pi}}{8.8\times 10^{10}~\rm{GeV}}\bigg)^2 \bigg( \frac{m_{\pi^a}}{1~\rm{eV}}\bigg)^{\frac12} \bigg(\frac{\theta_{a}}{\pi}\bigg)^2.
\ee
Demanding that the light dark pions made all the dark matter today gives
\bea
 F_{\pi} \geq 8.8\times 10^{9} {\rm{GeV}}.
\eea
Therefore, our dark sector with light pions has heavy baryons with mass,
\be
m_{b} \gtrsim 10^{10}~\rm{GeV}.
\ee
Moreover, the mass and millicharge of the quarks can be on one of the following ranges:
$$1.1\times 10^{-20} {\rm{eV}}\leq m_{\chi} \leq 1.1\times 10^{-11} {\rm{eV}} \quad \textmd{and} \quad \varepsilon e < \frac{1~ {\rm eV}}{F_{\pi}},$$
which put an extremely tight upper bound on the charge as $\varepsilon e < 10^{-19}$, or 
$$ m_{\chi} \leq 1.1\times 10^{-20} {\rm{eV}} \quad \textmd{and} \quad  \frac{1~ {\rm eV}}{F_{\pi}} < \varepsilon e < \frac{10^{4}~ {\rm eV}}{F_{\pi}},$$
which demands $\varepsilon e < 10^{-14}$.

\subsection{Baryon WIMPzillas}\label{FI-B}

The Dark QCD we consider also allows for stable dark baryons. These may be produced gravitationally at the end of inflation, or else through freeze-in production via their interactions with the SM. Here we consider and study three different freeze-in mechanisms, i.e. inflaton portal freeze-in, Higgs portal freeze-in, and millicharged QED freeze-in. We summarize these different possibilities in Fig. \ref{fig:mass-spectrum-I}.  Gravitational production of the WIMPzillas is shown in Fig.~\ref{fig:GPP}.

At this point we need to further specify the thermal evolution of the universe and in particular the reheating phase that follows cosmic inflation.  For the sake of generality, we consider the following phenomenological reheating model 
\be
\rho_{\rm reh} = \delta_{\rm reh} \bigg( \frac{a_{\rm inf}}{a_{\rm reh}}\bigg)^4 \rho_{\rm inf},
\ee
where $\rho_{\rm inf}=3M_{pl}^2H^2_{\rm inf}$ is the inflation energy density, $\delta_{\rm reh} \approx \big( a_{\rm reh}/a_{\rm inf}\big)^{-(3w-1)}$ is the efficiency of the reheating process, and $w$ is the effective equation of state in the intermediate period between the end of
inflation and the formation of the thermal bath. In case that the inflaton oscillates coherently about the minimum of the potential, its energy density redshifts as matter, i.e. $w=0$ with a $\delta_{\rm ref}>1$. On the other hand, if inflation ends with domination of the kinetic term, i.e. $w=1$, we have $\delta_{\rm ref}<1$. The reheating energy density is $\rho_{\rm reh}= \frac{\pi^2}{30} g_{*} T^4_{\rm reh}$. The temperature after reheating scales as $T\propto 1/a$, while between the end of inflation until reheating it scales differently and at the beginning it scales as   \cite{Kolb:2017jvz}
\begin{equation}
    T(a) \approx \big(\frac{a_{\rm inf}}{a}\big)^{\frac38(1+w)} T_{\rm max}
\end{equation}
where $a_{inf} < a< a_{\rm reh}$ and $T_{\rm max}$ is
\be
T_{\rm max} \approx 0.2  \bigg(\frac{100}{g_{eff}}\bigg)^{\frac18}  \rho^{\frac18}_{\rm inf}T_{\rm reh}^{\frac12}.
\ee

\subsubsection{Higgs portal}
The dark baryons can be coupled to the SM Higgs as
\be\label{H-b}
\mathcal{L}_{{\bf \rm{H}}} = \frac{y}{\Lambda_{{ \rm H}}}  {{\bf \rm H}}^{\dag} {{\bf \rm H}} ~ \sum_{i=p,n} \bar{b_i}b_i,
\ee
where we approximately take $m_p\approx m_n$ and $y=y_b \approx y_n $. This interaction can arise either from SM Higgs-dark quark interaction ${\cal L} \propto |H|^2\bar{\chi}\chi$, or  by dark-baryon-philic interactions, e.g,~ in a scenario with gauged baryon number \cite{Duerr:2013dza}. The latter case leaves the pion mass unchanged and only contributes to the baryon mass. In contrast, in the former case, after the electroweak phase transition, the coupling Higgs makes generates an effective mass for the dark pions as
$m_{\pi^0} \sim  246  \big(\frac{y\Lambda_x}{\Lambda_H}\big)^{\frac12} {\rm GeV}$, which, in the region of interest for freeze-in production, generically lifts the ULPs out of the ultra-light regime. For simplicity, here we treat the pion mass independently from the Higgs-baryon coupling.

At temperatures above the Electroweak (EW) symmetry breaking scale and $\Lambda_{x}> T_{\rm reh}$, the thermally averaged annihilation cross section of dark fermions with mass $m_b$ is  $\langle \sigma_{\rm H} v\rangle \approx \frac{1}{8\pi} \frac{y^2}{\Lambda^2_{\rm H}} \frac{3T}{m_b}$ \cite{Kolb:2017jvz}. The decay rate of the dark baryons associated with the Higgs interaction is 
\be
\Gamma_{\rm H} = \frac{3 T}{(2\pi)^{\frac53}} \big(\frac{y T}{\Lambda_{\rm H}}\big)^2 \big( \frac{m_b}{T}\big)^{\frac12} e^{-m_b/T}.
\ee 
Demanding that the WIMPzillas are never in thermal equilibrium with the SM, we find
\be
y^{-1}\Lambda_{\rm H} \gg  \frac{1}{(2\pi)^{\frac32}} \bigg( \frac{m_{b}}{T_{\rm max}} \bigg)^{\frac14}   e^{\frac{m_{b}}{2T_{\rm max}}} \sqrt{M_{pl} T_{\rm max}}.
\ee
For typical values of $m_{b} \sim 10^{10}~\rm GeV$, $m_{b}/T_{\rm max}\sim 10$, and $y\sim 0.1$, it gives $\Lambda  \sim 10^{14}~\rm GeV$.

This interaction generates a relic density of dark baryons via the freeze-in mechanism. Details of the calculations are provided in App. \ref{Appx}, and here we only report the final results. Concretely,
the relic density of the dark baryons produced through the Higgs portal is  (see Eq. \eqref{Omega-Higgs})
\be\label{Omega-H}
 \Omega_{b'} h^2 \simeq \mathcal{A} \bigg( \frac{y m_{b}}{\Lambda_H}\bigg)^2    \frac{\exp\big[10(3- \alpha \beta^{\frac12})\big]}{\beta^{4/(1+w)-1/2}}, \quad
\ee
where $\alpha$ and $\beta$ are
\be
\alpha \equiv \frac{m_{b}}{ \sqrt{H_{\rm inf}M_{pl}}}  \an  \beta \equiv \frac{\sqrt{H_{\rm inf}M_{pl}}}{T_{\rm reh} } ,
\ee
which are both more than one and $\mathcal{A} =\frac{(10\pi^2)^{-w/(1+w)}}{2(1+w)}$. We have $\mathcal{A}=1/2$ for $w=0$ and $\mathcal{A}=3/2\times 10^{-2}$ for $w=1$.

The abundance of dark baryons as a function of model parameters is shown in Fig.~\ref{fig:FI}, in the plane of $m_{b}/T_{\rm reh}$ and $m_{b}/\sqrt{M_{\rm pl} H_{\inf}}$, with the value of coupling $y$ indicated by background color. Dashed lines indicate $1\%$ of the dark matter in dark baryons while solid lines indicate $100\%$. Regions below the dashed and above the solid lines correspond to intermediate fractions between $1-100\%$. From this one may appreciate that, given values for any two of these parameter combination, one may tune the dark baryon abundance by varying the third. We define the  fraction of dark matter that this process produces as
\be
f_{b'} \equiv \frac{\Omega_{b'}}{\Omega_{\rm DM}},
 \ee
which may easily range from $0$-$1$.  This is illustrated in Fig.~\ref{fig:FI}. 

\subsubsection{Inflaton portal} 

We now consider a direct coupling of the dark baryons to the inflaton field, as
\be\label{Inf-b}
\mathcal{L}_{{\bf \rm{inf}}} = \frac{\lambda}{\Lambda_{{ \rm inf}}}  \phi^2~ \sum_{i=p,n} \bar{b_i}b_i,
\ee
where we approximately take $\lambda=\lambda_b \approx \lambda_n $. We consider an inflaton potential dominated by a quadratic term after inflation
\bea
V(\phi) \simeq \frac12 m_{\phi}^2 \phi^2.
\eea
Demanding that the dark baryons are never in thermal equilibrium gives 
\be
\lambda^{-1}\Lambda_{\rm inf} \gg  \frac{1}{\sqrt{2}(2\pi)^{\frac32}} \bigg( \frac{m_{b}}{T_{\rm max}} \bigg)^{\frac14}   e^{\frac{m_{b}}{2T_{\rm max}}} \sqrt{M_{pl} T_{\rm max}}.
\ee
Assuming for simplicity that $m_b \gg T_{max} \gg m_{\phi}$, and following the same procedure as for the Higgs portal, we find 
\bea\label{Omega-inf}
\Omega_{b'} \approx  \frac12 \mathcal{A} \bigg( \frac{\lambda m_{b}}{\Lambda_{\rm inf}}\bigg)^2    \frac{\exp\big[10(3- \alpha \beta^{\frac12})\big]}{\beta^{4/(1+w)-1/2}},
\eea
where ${\cal A}$, $\alpha$, and $\beta$, are defined as in the Higgs portal case.
The fraction of dark matter that the inflaton-portal can produce is again $f_{b'} = [0,1]$, which is illustrated in Fig.~\ref{fig:FI}.

\subsubsection{Millicharged QED}

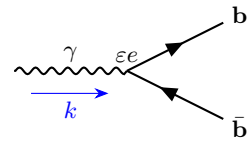
\begin{figure}
\centering
\begin{tikzpicture}[x=15mm, y=15mm]
\begin{feynman}
    \vertex[above=1] (a);
    \vertex[right=.15 of a] (b) ;
    \vertex[right=.15 of b] (c) ;
    \vertex at ($(c) + (1,0)$) (d) [label=\({\varepsilon e}\)];
    \vertex at ($(d) + (1, -0.5)$) (f4) {\(\bar{\bf{b}}\)};
    \vertex at ($(d) + (1,.5)$) (f5) {\(\bf{b}\)};
\diagram*{
    (c) -- [boson,  momentum'={[arrow style=blue]\(k\)}, edge label=\(\gamma\), thick] (d) -- [fermion, thick] (f5),
    (d) -- [anti fermion, thick] (f4),
};
\end{feynman}
\end{tikzpicture}
\caption{\label{Feyn} The millicharged QED (mQED) portal of the charged dark baryons. }
\end{figure}

The dominant electromagnetic processes between the dark baryons and electromagnetic plasma are
\be
\bar{f} f  \rightarrow  \bar{b}b \quad \textmd{and } \quad  \gamma^{*} \rightarrow  \bar{b}b,
\ee
which are proportional to $\varepsilon^2$, $f$ denotes SM charged fermions, and $\gamma^{*} $ denotes plasmons. The other possible process is $\gamma \gamma \rightarrow  \bar{b}b$ which is proportional to $\varepsilon^4$ and highly suppressed. The decay rate of dark baryons associated with the mQED processes is 
\bea
\Gamma_{em}  =  \frac{3 \varepsilon^2 \alpha^2_{em}}{(2\pi)^{\frac32}}  T \bigg( \frac{T}{m_b}\bigg)^{\frac32}  e^{-m_b/T}.
\eea
We demand that the WIMPzillas are never in thermal equilibrium with the SM. This puts an upper bound on the millicharge as
\be
\varepsilon \ll 4 \times 10^2  \big( \frac{m_b}{M_{pl}}\big)^{\frac12} \big( \frac{m_b}{T_{max}}\big)^{\frac14}  e^{\frac12 m_b/T_{max}}.
\ee
Given that $m_b \gtrsim 10^{10}~{\rm GeV}$ and $m_b> T_{max}$, this condition is satisfied throughout the parameter space for $\varepsilon \lesssim 10^{-2}$.

The relic density of the heavy dark baryons produced today through freeze-in by QED processes is  
\be\label{Omega-H}
 \Omega_{b'} h^2 \simeq  10^{-2}\mathcal{A} \varepsilon^2 \bigg( \frac{\varepsilon \alpha_{em}}{10^{-2}}\bigg)^2    \frac{\exp\big[10(3- \alpha \beta^{\frac12})\big]}{\beta^{4/(1+w)-1/2}}, \quad
\ee
where $\alpha_{em}=e^2/4\pi$, and $\mathcal{A}$, $\alpha$ and $\beta$ are similar to \eqref{Omega-H} (See Fig. \ref{fig:FI}). Fig. \ref{fig:FI} implies that generating significant fraction of dark baryons from the mQED interactions requires $\varepsilon \gtrsim 10^{-4}$. As we discuss later in Sec.~\ref{sec:constraints}, this possibility is ruled out by constraints on the millicharge the light pions.

\subsubsection{Gravitational production}

Finally, dark baryons may be produced gravitationally, without relying on any direct coupling to the Standard Model or the inflaton. So-called ``gravitational particle production'' \cite{Chung:1998rq,Chung:1998ua,Kuzmin:1998kk,Chung:2001cb,Redi:2020ffc} is generated by the non-adiabatic expansion of spacetime that occurs at the end of inflation. See \cite{Carney:2022gse} for a recent review.

\begin{figure}
    \centering
    \includegraphics[width=0.49\textwidth]{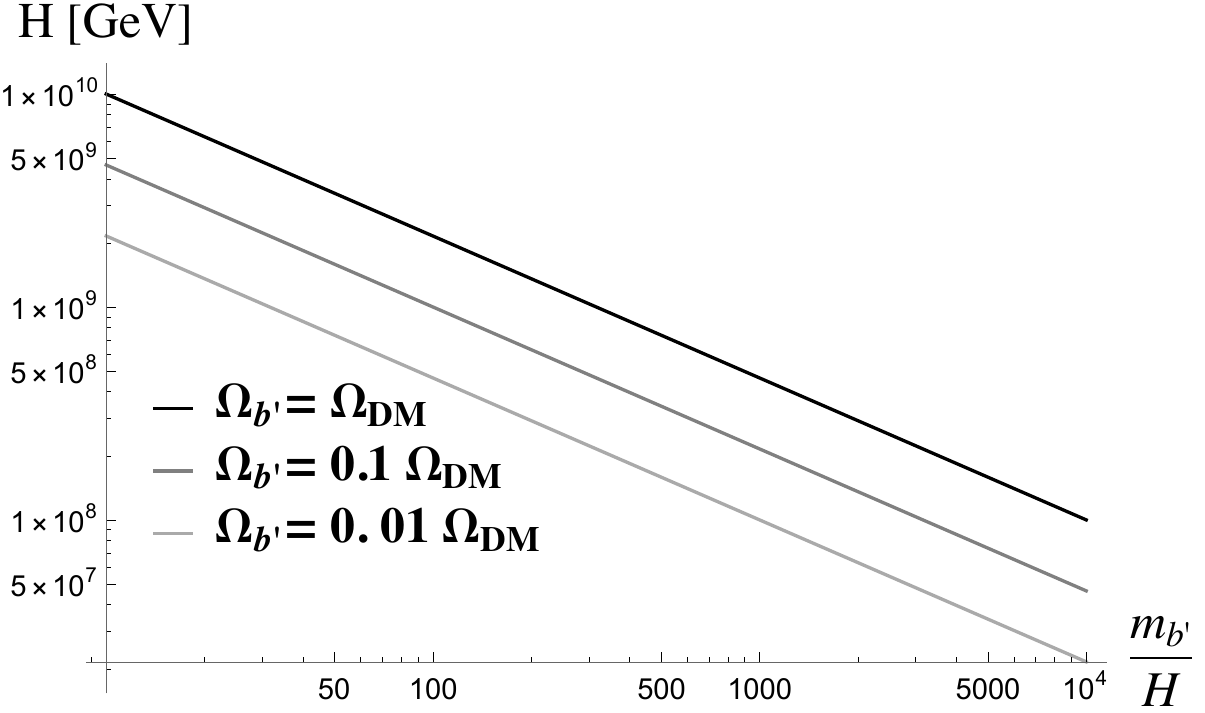}
    \caption{\label{fig:GPP} Gravitational Production of Dark Baryons WIMPzillas. The black, gray, and light gray, lines correspond to 100\%, 10\%, and 1\% of the observed dark matter in gravitationally produced dark baryons.}
\end{figure}

In our work we assume that the dark QCD theory is in the confining phase during inflation and reheating, and hence $m_{b} \gg H$. Gravitational production in this parameter regime has been extensively studied in recent years \cite{Chung:2018ayg,Ema:2015dka,Ema:2016hlw,Ema:2018ucl,Ema:2019yrd}. In particular, the dark matter density in the regime $H \ll m_b$, with $m_b \lesssim m_{\phi}$, with $m_{\phi}$ the mass of the inflaton, is given by \cite{Ema:2019yrd}
\begin{eqnarray}
	\frac{\rho_{b}}{s} \simeq &&4\times 10^{-10}{\rm GeV}\,\mathcal C  
	\left( \frac{m_{b}}{10^9\,{\rm GeV}} \right)\left( \frac{H}{10^9\,{\rm GeV}} \right) \\
	&& \times \left( \frac{T_{\rm re}}{10^{10}\,{\rm GeV}} \right)
	\left( \frac{m_{b}}{m_{\rm \phi}} \right)^2.
\end{eqnarray}
where ${\cal C} = 10^{-2} - 10^{-3}$ is a numerical constant, and $s$ is the entropy density. This may be compared to the observed DM abundance $\rho_{\rm DM}/s \sim 4 \times 10^{-10}$ GeV. As a fiducial example, we fix $m_\phi=10^{13}$ GeV and $T_{\rm re}=10^{11} {\rm GeV}$. The dark matter abundance, as a function of $H$ and the ratio $m_{b'}/H$ is shown in Fig.~\ref{fig:GPP}. In this figure the black line denotes parameters such that $\Omega_{\rm b'}$ gives $100\%$ of the observed dark matter abundance, while the gray and light gray lines correspond to $10\%$ and $1\%$ of the dark matter respectively.

\section{ULP Halos and Boson Stars}


\label{sec:halos}

We now consider cosmological and astrophysical implications of our model.
A striking feature of ultra-light bosonic dark matter candidates is the existence of self-gravitating soliton solutions. In the context of fuzzy dark matter, with $m \sim 10^{-22}$ eV, these solitons are on cosmological scales, forming the core of dark matter halos. For larger masses, such as the benchmark QCD axion mass $m\sim10^{-5}$ eV, the soliton solutions instead correspond to boson stars, which have their own signatures, such as gravitational waves. In both cases, the soliton solutions can be described in terms of wavefuntion, corresponding to the non-relativistic limit of the scalar field. The resulting equations of motion are Schrodinger and Poisson equations.  For a discussion of the allowed mass range and constraints, see \cite{Bar:2021kti}.

We seek to construct such soliton solutions in the context of ULP  dark matter.  To do so, we first decompose the charged pions $\pi_{\pm}$ into real components, as
\begin{equation}
\pi^{\pm} = \pi^1 \pm  i \pi^2
\end{equation}
and take the non-relativistic limit of $\pi^{0,1,2}$ as
\begin{equation}
\pi^{i} = \frac{1}{\sqrt{2m_i}}\left[ \psi_{i} e^{-i m_i t} + {\psi_i} ^* e^{i m_i t} \right],
\end{equation}
with $i=0,1,2$. The equations of motion of the system are then given by 3 copies of the Schr\"{o}dinger equation,
\begin{equation}
i \hbar \frac{\partial \psi_i}{\partial t} = - \frac{\hbar^2}{2m_i}\nabla^2 \psi_i + m_i \Phi \psi_i
\end{equation}
coupled via the Poisson equation,
\begin{equation}
\nabla^2 \Phi = 4 \pi G \displaystyle \sum_{i=0,1,2} m_i |\psi_i|^2 .
\end{equation}
where $m_{1,2}\equiv m_{\pm}$. For simplicity we have neglected non-gravitational interactions of the pions, which are suppressed by the pion decay constant. We leave this interesting aspect to future work.

\begin{figure*}
\includegraphics[width=0.49\textwidth]{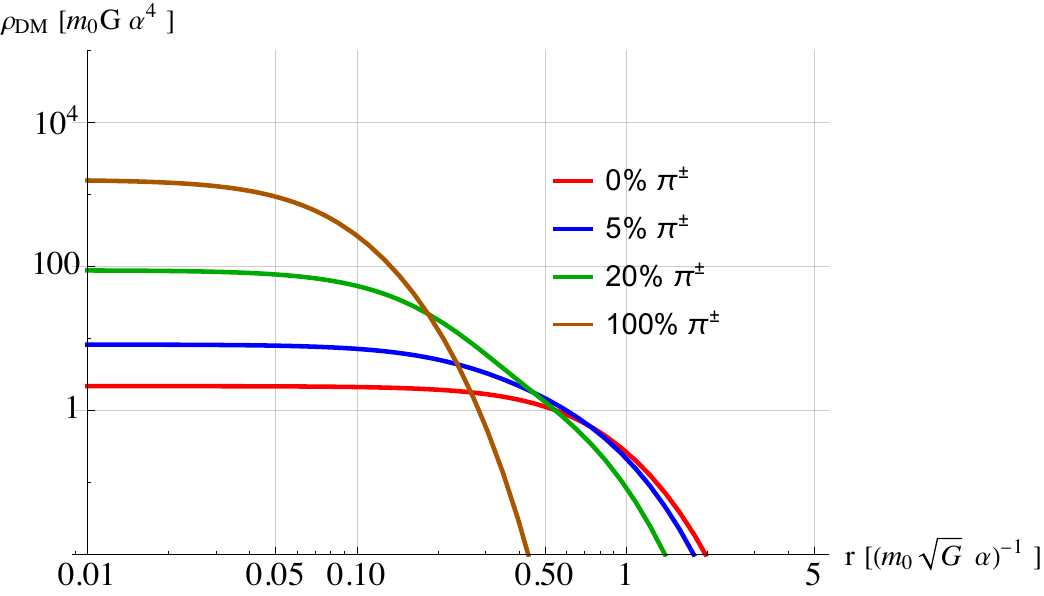}
\includegraphics[width=0.49\textwidth]{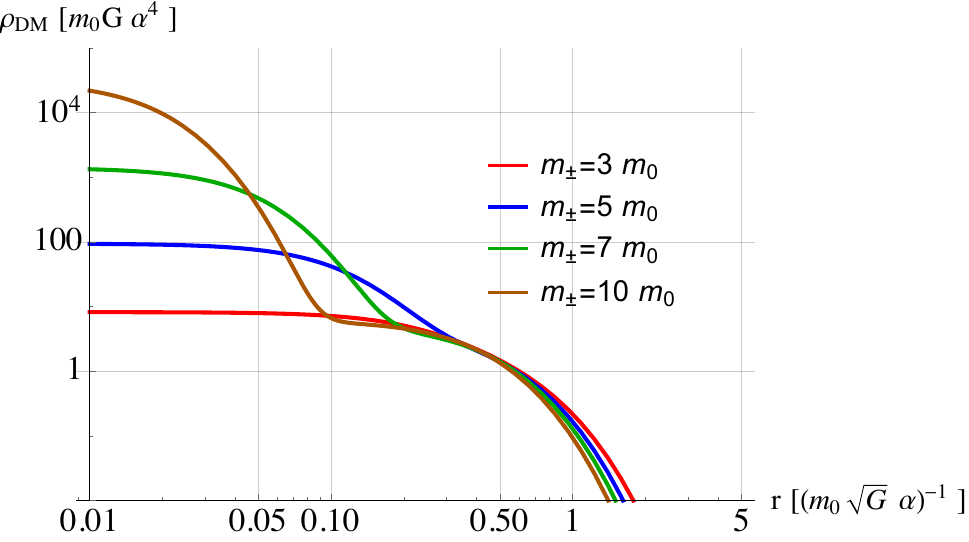}
\caption{\label{soliton} Soliton solutions to self-gravitating ULPs in the non-relativistic limit. All solutions have the same total mass, $M \equiv \sum_{i} m_i \int {\rm d}^3r  |\psi_i(r)|^2$. The units are defined with respect to an arbitrary energy scale $\alpha$ (when self-interactions are included this scale is fixed to be the decay constant  \cite{Schiappacasse:2017ham}). [{\it Left Panel}]  We fix mass ratio $m_{\pm}/m_0=3$ and vary the fraction of the mass of the soliton comprised of $\pi_{\pm}$. As the fraction increases, the soliton contracts and the central density grows larger. [{\it Right Panel}]  We fix the mass fraction $f_{\pm}=5\%$ and vary the mass ratio $m_{\pm}/m_0$. As the mass ratio grows larger, a distinct high-density inner soliton emerges at small radii. In all cases we have fixed the total mass of the halo.}
\end{figure*}

To construct soliton solutions, we follow the references \cite{Schiappacasse:2017ham}, \cite{Guo:2020tla}, and \cite{Eby:2020eas}. Assuming an ansatz for the  spatial dependence of the pion wavefunctions \cite{Schiappacasse:2017ham}
\begin{equation}
\psi_i(x) = \sqrt{\frac{3 N_i}{\pi^3 R_i^3}} {\rm sech} \left(\frac{r}{R_i} \right)
\end{equation}
the Hamiltonian of the system may be straight-forwardly derived as  \cite{Eby:2020eas}
\begin{equation}
H = \displaystyle \sum_{i=1} ^3 \left[ a \frac{N_i}{m_i R_i^2} + b \frac{m_i^2 N_i^2}{R_i} \right] +  \sqrt{2} b \displaystyle \sum_{\substack{i,j=1\\ i\neq j}} ^3\frac{m_{i}m_j N_{i}N_j}{\sqrt{R_i^2 + R_j ^2}}
\end{equation}
which generalizes the expression of \cite{Eby:2020eas} from two to three axions, but specializes to the  sech ansatz of \cite{Schiappacasse:2017ham}. The constants $a$ and $b$ are ansatz dependent, and for the sech ansatz are given by  \cite{Schiappacasse:2017ham}
\begin{equation}
a= \frac{12 + \pi^2}{6 \pi^2} \;\; , \;\; b= \frac{6}{\pi^4}\left( 12 \zeta(3) - \pi^2\right) .
\end{equation}
Finally, following  \cite{Schiappacasse:2017ham} we work in dimensionless variables, defined by the rescalings given in \cite{Schiappacasse:2017ham}.

We note that a general solution for the pion wavefunctions will correspond to a soliton with a net electric charge. For simplicity, we focus on solutions that are electrically neutral. The charge of a soliton may be expressed as,
\begin{equation}
Q = 4  \epsilon e   \int {\rm d}^3 x \left( \dot{\pi}_1 \pi_2 - \dot{\pi}_2 \pi_1  \right),
\end{equation}
where we have decomposed the standard relation for a charged complex scalar field into the real scalar components. An electrically neutral soliton corresponds to one of three cases: $\pi_1=\pi_2$, $\pi_1=0$, or $\pi_2=0$. We consider the case $\pi_1 = \pi_2$, and hence $\psi_1=\psi_2$.

To find an approximate ground state solution, building on the procedure of \cite{Eby:2020eas} , we vary the Hamiltonian with respect the radii $R_i$, while holding fixed both the total mass of the halo and the relative fraction of the mass contained within each dark matter component. We fix the total mass,
\begin{equation}
M_{\rm tot} =  \displaystyle \sum_{i=1} ^3  M_i ,
\end{equation}
where
\begin{equation}
M_i \equiv m_i N_i = m_i \int {\rm d}^3 r |\psi_i(r)|^2 .
\end{equation}
We also fix the relative fraction of the mass in each component, 
\begin{equation}
f_{i} = \frac{N_i}{N_{\rm tot}} .
\end{equation}
One may reasonably expect that $f_i$ depends on local environment of dark matter solitons, and on the primordial abundances of the ULPs. In this sense, varying $f_i$ corresponds to accounting for the diversity of dark matter halos.

Soliton solutions for varying $f_{i}$ are given in the left panel of Fig.~\ref{soliton}, where we consider the simple case that $m_{\pm}=3 m_0$. In all cases we explicitly confirm that the Hessian of the Hamiltonian is positive definite, and thus the solutions are stable. We consider four fiducial example solitons, corresponding a purely $\pi^0$ halo, and halos with $5 \%$, $20\%$ , and $100\%$ of the mass contained in $\pi^\pm$. One may easily appreciate that by adjusting the fraction of dark matter in the charged vs. neutral pions, the central density becomes higher and the core radius smaller. This effect becomes more dramatic as the ratio of pion masses is made larger. While we restrict ourselves here to a mild mass hierarchy, we note this is made partly for illustrative purposes, and additionally so that the non-relativistic limit may be consistently applied to both fields. 

The impact of raising the charged pion mass is shown in the right panel of Fig.~\ref{soliton}, where we fix the charged pions to be 5\% of the mass of the halo, and adjust the charged pion mass. For larger values of the charged pion mass, a distinct soliton is present at small radii, with a density that is orders of magnitude larger than the outer edges of the density profile. 

The diversity of ULP soliton density profiles is particularly interesting given the mild tension between the diversity of observed halos \cite{Oman:2015xda} and the predicted universal properties of Fuzzy dark matter solitons, see, e.g., \cite{Bar:2021kti,Lelli:2016zqa,Chan:2021bja} for a discussion and constraints. In the ULP model, a diversity of halos naturally emerges from adjusting the mixture of pions that make up the dark matter halo. In the case of a large high hierarchy, e.g. if $\pi^{0}$ is in the fuzzy range while $\pi^{\pm}$ is in the conventional QCD axion range, dark matter halos can be expected to range from an the Navarro–Frenk–White (NFW) profile to the cored halos familiar from fuzzy dark matter.


\section{Current Constraints and Avenues for Detection}

\label{sec:constraints}

\begin{table*}
\begin{center}
\begin{tabular}{|l|l|l|l|l|}
\hline
 & ~~~~~Method  & ~~Constraint on $\varepsilon$ & Reference & ~~~Avoidance \\
\hline
\multirow{5}{*}{\begin{sideways} Astrophysics \end{sideways}} &  Supernova cooling & $10^{-9} < \varepsilon < 10^{-7}$ & \cite{Davidson:2000hf} & \\
& Stellar cooling & $\varepsilon < 1.7\times10^{-14}$ & \cite{Davidson:2000hf,Raffelt:1996wa} & \\
& Solar cooling & $\varepsilon < 10^{-13.6}$ & \cite{Vinyoles:2015khy} & \\
& Magnetars & $\varepsilon^2(\frac{m}{\text{eV}}) < 10^{-16}$ & \cite{Korwar:2017dio} & \\
& Milky Way satellites & $\varepsilon < 10^{-15} (\frac{m}{\text{eV}})$ & \cite{Nadler:2019zrb, Bogorad:2021uew} & $f_{\rm \pi^{\pm}} < 10 \%$\\
\hline 
\multirow{6}{*}{\begin{sideways} Cosmology \end{sideways}} & BBN & $\varepsilon < 2.1\times10^{-9}$ & \cite{Davidson:1993sj,Davidson:2000hf,Vogel:2013raa} & \\
& CMB & $\epsilon \lesssim 2 \times 10^{-12} \big(\frac{m}{{\rm{keV}} }\big)$ & \cite{Berlin:2019uco, Dvorkin:2019zdi,Boddy:2018wzy} &  $f_{\pi^{\pm}} < 0.4\% $\\
& SZ effect & $\varepsilon < 2\times10^{-9}$ & \cite{Burrage:2009yz} & $m_{\pi^{\pm}} \gg 10^{-7}$ eV\\
& SN dimming & $\varepsilon < 4\times10^{-9}$ & \cite{Ahlers:2007qf} & $m_{\pi^{\pm}} \gg 10^{-7}$ eV\\
& Pulsar timing and FRBs & $\varepsilon(\frac{\text{eV}}{m}) < 10^{-8}$ & \cite{Caputo:2019tms} & $f_{\rm \pi^{\pm}} < 10 \%$\\ 
\hline
\multirow{5}{*}{\begin{sideways} Laboratory \end{sideways}} & Laser experiments & $\varepsilon < 3\times10^{-6}$ & \cite{Ahlers:2007qf} &\\
& Lamb shift & $\varepsilon < 10^{-4}$ & \cite{Gluck:2007ia} &\\
& Positronium & $\varepsilon <  3.4\times10^{-5}$  & \cite{Badertscher:2006fm} &\\
& Coulomb's law deviations & $\varepsilon \lesssim5\times10^{-6}$ & \cite{Jaeckel:2009dh} & $m_{\pi^{\pm}} \gg 1$ eV\\
& Schwinger effect in cavities & $\varepsilon \lesssim 10^{-6}$ & \cite{Gies:2006hv} & $m_{\pi^{\pm}} \gg 1$ eV\\
\hline
\end{tabular}
\end{center}
\caption{ The current (astrophysical, cosmological, and laboratory) constraints on MCPs with $m_{_{\rm MCP}} \lesssim 1$ MeV, such as light and ultra-light pions. (For possible proposed mechanisms to evade stellar bounds see \cite{DeRocco:2020xdt} and the references therein.) The column ``avoidance'' refers to the region of parameter space where the constraint of that row does not apply. Here $f_{\pi}$ denotes the fraction of dark matter, not the decay constant. }
\label{Table2}
\end{table*}

In this section, we discuss the current constraints and the possibility of detection of the millicharged particles  (MCP), as well as the dark QCD phase transition, and ULPs more generally.

\textit{Bounds on millicharge: }  The constraints on light MCPs are very strong.  In Table \ref{Table2} we summarize the current astrophysical, cosmological, and laboratory bounds on MCPs with $m_{_{\rm MCP}} \lesssim 1$ MeV. Some of these constraints can be avoided in our setup, as indicated in the last column of the table.  The strongest bound on MCP dark matter comes from CMB, leading to the constraint \cite{Berlin:2019uco, Dvorkin:2019zdi,Boddy:2018wzy}
\be
\varepsilon < 10^{-15}\bigg(\frac{m}{\rm{eV}}\bigg).
\ee
This constraint can be avoided in our setup once the fraction of the DM density in the charged pions is less than 0.4\%, i.e., $f_{\pi^{\pm}} <0.1$, or if the charged pion mass is in the ultralight regime (it is noteworthy to mention that CMB constraints on millicharged coherent scalars, e.g., $m< 1$ eV, have never been worked out, and we defer to future work.) Note this still leaves open the possibility of ULPs comprising 100\% of the dark matter, with $f_{\pi^0} \gtrsim 0.996$. As a result, the allowed parameter space for our charged pions is greatly enlarged compared to standard scenario of MCP dark matter. The strongest bound on the millicharge, that cannot be avoided, comes from the stellar cooling that gives \cite{Davidson:2000hf,Raffelt:1996wa}
\be
\varepsilon < 1.7 \times 10^{-14}.
\ee
Several scenarios are proposed in the literature to evade the stellar bounds. For details see \cite{DeRocco:2020xdt} and the references therein. The millicharged dark pions also interact with the SM baryons. The Rutherford-type scattering
cross-section of the pions off of a SM proton through a photon is
\bea
\sigma_{\pi^{\pm}b} \sim \frac{\alpha^2_{em} \varepsilon^2}{\mu^2v_{rel}^4 },
\eea
where $v_{rel}$ is its relative velocity, and $\mu\approx m_{\pi^{\pm}}$ is the SM proton and dark pions reduced mass. 
That gives an upper bound on the charge as $\varepsilon < 10^{-15} (\frac{m}{\text{eV}})$ \cite{Nadler:2019zrb, Bogorad:2021uew}. For detailed discussion of the phenomenology and detection possibility of generic light millicharged dark matter see \cite{Bogorad:2021uew}. 
The parameter space of our setup, therefore, is divided into two parts:
\begin{itemize}
    \item charged pions make more than 0.4\% of the dark matter with millicharge $\varepsilon<10^{-15}\big(\frac{m_{\pi^{\pm}}}{\rm{eV}}\big)
$,
\item  charge pions make a subdominant part of the dark matter, i.e.  $f_{\pi^{\pm}}<0.004$ , with millicharge $\varepsilon < 1.7 \times 10^{-14}$.
\end{itemize}

\textit{Detection of millicharge: } The main approach to detect light millicharged particles is the direct deflection method \cite{Berlin:2019uco}.  This experiment is based on distorting the local dark matter flow with time-varying fields and measuring these distortions with shielded resonant detectors. The expected reach of the direct deflection experiment is MCPs in the range $1-10^{7}~\rm{eV}$ and charge in the range $\varepsilon=10^{-16}-10^{-9}$ which can be improved by one order of magnitude in the long-term project \cite{Berlin:2019uco, Bogorad:2021uew}. This can be used to search for pions in the ``light'' range, i.e., $m_{\pi} = [1 {\rm \, eV}, {10 {\rm \, keV}}]$.

\textit{Detection of dark QCD phase transition: } If the dark QCD phase transition is first order, it can produce gravitational waves background with the frequency peak  $f \gtrsim 10^4 \rm{Hz}$. Such frequencies are above the LIGO/Virgo band and in the range of ultra high-frequency gravitational waves. For a recent review on several detector concepts which have been proposed to detect this ultra-high frequency signals see \cite{Aggarwal:2020olq}. 

\textit{Other directions for ULP discovery: } Analogous to axion electrodynamics, the ULP model exhibits {\it ULP-electrodynamics}. This is a described by the Lagrangian,
\begin{equation}
 {\cal L}_{\rm ULP-EM} = {\cal L}_{\pi} + \frac{1}{4}F^2 +  \epsilon^2 e^2\pi^+ \pi^- A_\mu A^\mu + g \frac{\pi^0}{4 F_{\pi}} F \tilde{F}   ,
\end{equation}
where ${\cal L}_{\pi}$ corresponds to the mass and kinetic terms of the ULPs. From this one may derive the modified vacuum Maxwell equations as,
\begin{eqnarray}
\vec{\nabla} \cdot \vec{E} && = \frac{g}{F_{\pi}} \vec{B} \cdot \vec{\nabla} \pi^0 - \epsilon^2e^2\pi^+ \pi^- V  \\
\vec{\nabla}\times \vec{B} - \frac{\partial E}{\partial t} && = \frac{g}{F_{\pi}} \left(  \vec{E} \times \vec{\nabla} \pi^0 - \vec{B} \frac{\partial \pi^0}{\partial t} \right) - \epsilon^2e^2 \pi^+ \pi^- \vec{A} .\nonumber
\end{eqnarray}
We defer to future work an exploration of the ULP signal for axion detection experiments such as ADMX \cite{ADMX:2020ote}. One might also search for these interactions in form of ``Cosmic ULP backgrounds'', analogous to the recently proposed Cosmic Axion Background \cite{Dror:2021nyr}, but with one cosmic background for each ULP. Also in a cosmological context, the coupling of photons to the charged pions may be an additional source of resonant production of the former. The equation of motion for a Fourier mode of the photon field $A^\mu$, with wavenumber $k$ and polarization $\pm$, in an Friedmann–Lema\^{i}tre–Robertson–Walker  (FLRW) spacetime, is given by,
\begin{equation}
    A_{k\pm } '' + (k^2 + \epsilon^2 e^2 \pi^+ \pi^- \pm \frac{g}{ F_{\pi}}k {\pi^{0}}\,' )A_{k\pm }=0,
\end{equation}
where $'$ denotes a derivative with respect to conformal time. This above differs from the more conventional axion case (see e.g. \cite{Finelli:2008jv}) by the mass-like coupling to the charged pions. The latter provides an additional mechanism for parametric resonance production of photons, as has been studied in an axion context in e.g. \cite{Hertzberg:2018zte}. 

\section{Discussion} \label{sec:discussion}


In this work we have developed the theory of Ultra-Light Pion (ULP) dark matter and dark baryon WIMPZillas, which together comprise the ULP-WIMPzilla model. In this model, ultra-light dark matter arises as composite states of a confining gauge theory, namely the Goldstone bosons of chiral symmetry breaking, analogous to the Standard Model pions. In the limit of very small dark quark masses, $m_\chi \lesssim 10^{-19} {\rm eV}$, and high confinement scale $\Lambda_x \gtrsim 10^{10} {\rm GeV}$, the dark pions enjoy an axion-like cosmological history and can provide the observed abundance of dark matter. The mass spectrum of pions encodes the charge and confinement scale of the dark QCD-like theory, and is in turn encoded in the density profile of dark matter halos (or boson stars, depending on the mass of the pions).

As the name would suggest, the pions themselves are only part of the ULP-WIMPzilla model. There are additional degrees of freedom in the theory which may exhibit interesting dynamics, such as the dark baryons. Due to the high confinement scale, the dark baryons naturally realize the WIMPZilla paradigm. Produced either gravitationally or via freeze-in, the dark baryons can constitute a small to significant fraction of the dark matter for a wide range of parameters.

The are a few things worth highlighting before closing this article:
\begin{enumerate}
    \item The ULP-WIMPzilla model is the first scenario of ultra-light ($m < {\rm eV}$) dark matter in a confining gauge theory, and the first example of an electrically millicharged ultra-light ($m < {\rm eV}$) dark matter candidate. The related but distinct mass range $m = [{\rm eV}, 10 \, {\rm keV}]$ is studied in \cite{Bogorad:2021uew}. 
    \item ULPs are the first model of 3 ultralight scalars, building on previous work on two ultralight scalars \cite{Guo:2020tla,Eby:2020eas}. The ULP model predicts two of the three scalars are degenerate in mass, and predicts predicts a mass splitting with the third set by the millicharge. 
    \item The ULP-WIMPzilla model is the first scenario to unify ultralight and fermionic WIMPZilla dark matter, with the two components unavoidably connected by common underlying parameters. Dark matter is generically an admixture of neutral pions, charged pions, and baryon WIMPzillas. This leads to a diversity of dark matter halos.
    \item Depending on the millicharge, the quark mass, and confinement scale, the charged pions may have a mass comparable to the neutral pion or may be much heavier. In the latter case, the neutral pion can be wave-like (or `fuzzy') while the charged pions may be wave-like or particle-like.  Charged pions with $m_{\pi^\pm} > 10 \, {\rm keV}$, exhibit their own phenomenology, which we defer to future work.
    \item The strongest constraint on the light millicharge by CMB, is alleviated if the charged pions are a subdominant component of the dark matter, $f_{\pi^\pm}< 0.4\%$. This is independent of the fraction of DM in the neutral pions or dark baryons. This significantly opens up the parameter space of the model, while being testable at future experiments.  The ultra-light scalar with $m< 1$ eV behaves as a coherent state and the CMB constraints on that case have never been worked out. We leave this effect to future work.
\end{enumerate}

Finally, we address an important question: how can ULPs, and their UV completion in dark QCD, be distinguished experimentally from a conventional axion model, with a UV completion in scalar fields, and which is devoid of strong interactions? The suggestion presented in this work is to perform a dedicated search for the whole structure of the theory, namely, the ULP-WIMPzilla model, its pattern of charges and masses, and the associated phenomenology (such as gravitational waves from the dark QCD phase transition). This proposal sidesteps a related but distinct question: Given a detection at ADMX or related axion-search -- how may we test the nature of the axion as fundamental or composite? We leave this interesting question to future work.

\vspace{1cm}
{\bf Acknowledgements}

The authors thank Cora Dvorkin, Wayne Hu, Eiichiro Komatsu, Joachim Kopp, and Matthew McCullough, for discussions and comments. EM is supported in part by a Discovery Grant from the National Science and Engineering Research Council of Canada. 

\appendix

\section{Number density of heavy dark baryons}\label{Appx}

This appendix presents the calculations of heavy fermion generation by a Higgs portal of the form Eq. \eqref{H-b}. Here we followed the analysis performed in \cite{Kolb:2017jvz}.
At temperatures above the EW symmetry breaking ($T> 100~GeV$), the Higgs field includes two states i.e. ${\bf \rm{H}} = \begin{pmatrix} \rm H^+ \\ \rm H^{0} \end{pmatrix}$, and hence there are two annihilation channels to dark baryons 
\be
{\rm H^0  H^0} \rightarrow \bar{\boldsymbol{b}}_{x} \boldsymbol{b}_x \an {\rm H^+ H^-} \rightarrow \bar{\boldsymbol{b}}_x \boldsymbol{b}_x.
\ee
Due to the isospin symmetry, the matrix elements of these two channels are equivalent, and the final
thermally averaged cross section is doubled
The thermally averaged annihilation cross section is 
\bea\label{Higgs}
\langle \sigma_{\rm H} v \rangle &=& \frac{4^2}{n_{\rm eq}n_{\rm eq}} \int \frac{d^3\vec{p}_1}{(2\pi)^3} \int \frac{d^3\vec{p}_2}{(2\pi)^3} ~ \bar{\sigma}_{{\rm H^{\dag} H}\rightarrow \bar{b}_xb_x} ~ v_{ \textmd{M\o l}}(p_1,p_2) \nonumber\\
&\times & \exp(-(E_1+E_2)/T),
\eea
where $\bar{\sigma}_{{\rm H^{\dag} H}\rightarrow \bar{b}_x b_x}$ is the spin-averaged annihilation cross section and $v_{\textmd{M\o l}}(p_1,p_2)$ is the M\o ller velocity
\be
v_{\textmd{M\o l}}(p_1,p_2) = \sqrt{ \vert \vec{v}_1-\vec{v}_2 \vert^2 - \vert  \vec{v}_1 \times  \vec{v}_2\vert},
\ee
here $\vec{v} = \vec{P}/E$. Since $\bar{\sigma}_{{\rm H^{\dag}\rm H}\rightarrow \bar{b}_x b_x}$ is a function of $s$ only, Eq. \eqref{Higgs} can be further simplified as \cite{Kolb:2017jvz}
\bea\label{Higgs}
\langle \sigma_{\rm H} v \rangle &=& \frac{4^2}{n_{\rm eq}n_{\rm eq}} \frac{T}{32\pi^4} \int^{\infty}_{4m^2_{b}}  ~ ds \sqrt{s} (s-4m^2_{b}) K_1\big(\frac{\sqrt{s}}{T}\big) \nonumber\\
&\times& \bar{\sigma}_{{\rm H^{\dag}\rm H}\rightarrow \bar{b}_x b_x}(s) ,
\eea
where $s=(p_1+p_2)^2$, $K_1$ is the modified Bessel function of the second kind of order 1, and $\bar{\sigma}_{{\rm H^{\dag}\rm H} \rightarrow \bar{b}_x b_x}$ is the cross section associated to the Higgs annihilation 
\be
\bar{\sigma}_{{\rm H^{\dag}\rm H}\rightarrow \bar{b}_x b_x}(s)   =   \frac{1}{32\pi}\frac{\lambda^2}{\Lambda^2 _{\bf \rm{H}}} \frac{1}{s} \sqrt{s-4m^2_{b}} \sqrt{s-4m^2 _{{\bf \rm{H}}}}.
\ee
In the limit that $m_{b}\gg T$, the thermally averaged annihilation cross section of dark fermions with mass $m$ is \cite{Kolb:2017jvz}
\be
\langle \sigma_{\rm H} v\rangle \approx \frac{1}{8\pi} \frac{\lambda^2}{\Lambda^2_{\rm H}} \frac{3T}{m_{b}} .
\ee
The number density of dark baryons generated by the SM Higgs is
\bea
&& n_{b}(t) = a^{-3}(t) \int_{a_{\rm inf}}^{a(t)} \frac{d\ln a}{H}~ a^3 \langle \Gamma_{\rm H}\rangle n_{\rm eq}  \nonumber\\
&\approx &  \frac{4/(2\pi)^4}{(1+w)} \bigg( \frac{a_{inf}}{a(t)} \bigg)^3 \bigg( \frac{\lambda }{\Lambda_H}\bigg)^2 \frac{ m_{b} T_{max}^{5}}{H_{inf}}  \exp[-\frac{2m_{b}}{T_{max}}], \nonumber\\
\eea
where $n_{\rm eq} = 4 \big(\frac{m_{b}T}{2\pi}\big)^{\frac32} e^{-\frac{m_{b}}{T}}$ we used $m_{b}\gg T_{reh}$ limit and we have
\be
\frac{a_{inf}}{a(t_0)} = \frac{a_{inf}}{a_{reh}}  \frac{a_{reh}}{a(t_0)} = \bigg(\frac{\rho_{reh}}{\rho_{inf}}\bigg)^{\frac{1}{3(1+w)}} \bigg(\frac{g_{0,s}}{g_{reh,s}}\bigg)^{\frac13} \frac{T_{0}}{T_{reh}}.
\ee
Here $T_0$ is the temperature of the Universe today and $g_{reh,s}$ and $g_{0,s}=3.91$ are the effective number of relativistic degrees of freedom contributing to the total entropy at reheating and today respectively.
The relic density of the dark baryons produced today through the Higgs portal is 
\bea
&& \Omega_{b'} \equiv \frac{m_{b}n_{b}(t_0)}{\rho_{crit}}  \nonumber\\
&\approx & \frac{5 \times 10^{10}/(2\pi)^4}{1+w} \frac{g_{0,s}}{g_{reh,s}} \bigg(\frac{T_{0}}{T_{reh}}\bigg)^3 \bigg( \frac{\lambda }{\Lambda_H}\bigg)^2  m^2_{b}   \nonumber\\
&\times& (0.2)^5 3^{\frac58} \frac{3.91}{100} \big( 10\pi^2\big)^{1/(1+w)} \bigg( \frac{g_{eff}(T_{reh})}{100}\bigg)^{1/(w+1) - 13/8}  \nonumber\\ &\times& \frac{T_{reh}^{4/(1+w)+5/2}}{(M_{pl}H_{inf})^{2/(1+w)-1/4}} M_{pl} \exp[-\frac{2m_{b}}{T_{max}}]  eV^{-4} , \nonumber
\eea
where $\rho_{crit} = 0.8 \times 10^{-10}~eV^4 $ is the total energy density of the universe today. In case that $T_{reh}\gg T_{EW}$ and using the fact that $T_0=0.24\times 10^{-3} eV$ and $M_{pl}=2.4 \times 10^{18} GeV$, we have
\bea\label{Omega-Higgs}
&& \Omega_{b'} \approx 
\nonumber\\
&& \frac{5.6 \times 10^{15}}{(2\pi)^{2}} \frac{(10\pi^2)^{-\frac{w}{(1+w)}}}{(1+w)} \bigg( \frac{\lambda m_{b}}{\Lambda_H}\bigg)^2  \bigg(\frac{T^2_{reh}}{M_{pl}H_{inf}}\bigg)^{\frac{2}{(1+w)}-\frac14}  \nonumber \\
&\times & \exp\bigg[-10 \bigg(\frac{m^2_{b}}{M_{pl}H_{inf}}\bigg)^{\!\frac12} \bigg(\frac{M_{pl}H_{inf}}{T^2_{reh}}\bigg)^{\!\frac14}\bigg].    
\eea

\bibliographystyle{JHEP}
\bibliography{ref}

\end{document}